\documentclass[a4paper,12pt]{article}

\title{Exactly soluble isotropic spin-1/2 ladder models}

\author{
V. Gritsev and D. Baeriswyl \\ 
{\it D\'epartement de Physique, Universit\'e de Fribourg},\\ 
{\it P\'erolles, CH-1700 Fribourg, Switzerland}
}

\date{}

\begin{document}

\maketitle

\abstract{
The undeformed limit of the dilute two-color braid-monoid algebra gives a natural basis for the description of spin-1/2 ladder models, and allows different Baxterization Ans\"{a}tze. Based on this observation we find the complete classification of exactly soluble generalized isotropic spin-1/2 ladder models.
}

\section{Introduction}
Exactly soluble models afford considerable insight into the complicated behavior of frustrated quasi-one-dimensional spin systems.
Here we investigate the  most general spin ladder which includes  isotropic  Heisenberg 
leg, rung and diagonal interactions, 
as well as biquadratic leg-leg, diagonal-diagonal and rung-rung spin exchange terms. The ladder system represents the minimal model admiting both  multiple-spin exchange interactions and non-trivial exact solutions.

The spin models on a ladder are usually characterized in terms of their symmetry, i.e., the invariance of the Hamiltonian with respect to transformations in ordinary space and in spin space. In Ref.~\cite{Wang} the $su(4)$ and $su(3|1)$- invariant integrable Hamiltonians have been constructed on the basis of the solution of the Yang-Baxter equation ($R$-matrix), given in terms of the corresponding permutation operator. Multiparametric models related to these symmetries have been considered in \cite{FLT} and \cite{TFHL} while an extension to $n$-leg ladder models associated with the $su(2^{n})$, $o(2^{n})$ and $sp(2^{n})$ algebras has been obtained in \cite{BM} and \cite{BGLM}. The $R$-matrix of the generalized ladder model \cite{BM,AFW} satisfy the Yang-Baxter equation with a multiplicative composition law.
Physical properties of some exactly soluble ladder systems, such as magnetisation plateaus and thermodynamics have been presented in \cite{Wang}, \cite{GB}, \cite{TDFM}, \cite{BGFT}, \cite{MBG}. 

From the general point of view, many known examples of solutions of the Yang-Baxter equation are  based on algebraic structures, such as braid algebra, Temperley-Lieb and Hecke algebras, Bir\-man-Wen\-zl-Mu\-ra\-ka\-mi algebra. Therefore it is natural to search for the algebraic construction suitable for describing $SU(2)$-invariant spin ladders.

We study exactly soluble models within the space of coupling parameters. The general ladder Hamiltonian is based on a plaquette algebra, 
by which is meant an algebra of operators defined on the four-spin 
plaquette units of the ladder. This algebra satisfies quasi-local 
commutation relations, meaning that operators defined on next-to-nearest (and 
further-) neighbor plaquettes commute. Using the commutation relations 
of this algebra we find exact expressions for different $R$-matrices 
which give solutions to the Yang-Baxter equation. These solutions 
correspond in turn to soluble (and $SU(2)$-invariant) spin-ladder 
Hamiltonians, which may be either gapless or gapped. 

The plaquette algebra provides a representation of the undeformed limit of a dilute, two-color braid-monoid algebra. The dilute, two-color braid-monoid, or two-color Bir\-man-Wen\-zl-Mu\-ra\-ka\-mi algebra was introduced and investigated in a series of papers by Grimm and coworkers \cite{Grimm1}-\cite{Grimm5}. In particular situations, when the generators of the undeformed  dilute algebra are combined into full-braid and monoid generators, the algebra is equivalent to the Brauer \cite{Brauer} or braid-monoid algebra \cite{WDA}, which is the undeformed limit of the Bir\-man-Wen\-zl-Mu\-ra\-ka\-mi algebra. By taking all possible different combinations of generators of the algebra we construct the solutions of the Yang-Baxter equation, and thus the corresponding exactly soluble Hamiltonians. Because the algebra provides a complete basis for $SU(2)$ invariant ladder we believe that the list of isotropic $SU(2)$-invariant integrable ladder models (with short-range interactions) presented here is complete.

In Section 2 we introduce the general spin-1/2 ladder model and express all isotropic interactions in the plaquette-algebra basis. By seeking commutation relations of this algebra we observe that it provides a representation of the undeformed limit of a dilute two-color braid-monoid algebra. In Section 3 we use the commutation relations of the algebra to establish a number of Baxterization Ans\"{a}tze by considering different subsets of the generators of the algebra. We thus obtain all exactly soluble Hamiltonians in the parameter space of the generalized ladder Hamiltonian. In Section 4 we point out the relation between words, defined on the plaquette algebra and Matrix-Product ansatz states. In the concluding section we summarize the results and make contact with known integrable models.

\section{The ladder model and its algebraic structure}

\subsection{Model and operator basis}

The Hamiltonian of the two-leg ladder under consideration is
\begin{eqnarray}
H= \sum_{i} J_{L}({\bf S}_{i}\cdot {\bf S}_{i+1} + {\bf T}_{i}\cdot {\bf T}_{i+1})+ J_{R}{\bf S}_{i}\cdot {\bf T}_{i} +
J_{D}({\bf S}_{i}\cdot {\bf T}_{i+1} + {\bf T}_{i}\cdot {\bf S}_{i+1})\nonumber\\
+V_{LL}({\bf S}_{i}\cdot {\bf S}_{i+1})({\bf T}_{i}\cdot {\bf T}_{i+1})
+V_{RR}({\bf S}_{i}\cdot {\bf T}_{i})({\bf S}_{i+1}\cdot {\bf T}_{i+1})+V_{DD}({\bf S}_{i}\cdot {\bf T}_{i+1})({\bf S}_{i+1}\cdot {\bf T}_{i}),
\label{ladham}
\end{eqnarray}
where we include leg, $(J_{L})$, rung, $(J_{R})$, and diagonal $(J_{D})$ Heisenberg interactions, as well as leg-leg $(V_{LL})$, rung-rung, $(V_{RR})$,  and diagonal-diagonal, $(V_{DD})$, four-spin interactions. Spin-1/2 operators on rung $i$ are represented respectively by ${\bf S}_{i}$ and ${\bf T}_i$, $i=1,\ldots ,L$, and periodic boundary conditions are assumed, ${\bf S}_{L+1}={\bf S}_{1}$ and ${\bf T}_{L+1}={\bf T}_{1}$. Many solutions listed in the next Section are invariant with respect to the intertwinning transformation 
\begin{eqnarray}
{\bf S}_{2i}\leftrightarrow {\bf S}_{2i} \  &  \ {\bf T}_{2i}\leftrightarrow {\bf T}_{2i} \ & \  {\bf T}_{2i+1}\leftrightarrow {\bf S}_{2i+1},\nonumber\\
& J_{L}\leftrightarrow J_{D} \ \ & \ \ V_{LL}\leftrightarrow V_{DD}
\label{inter}
\end{eqnarray}
which interchanges ${\bf S}$ and ${\bf T}$ on every second rung and therefore we present one representative Hamiltonian for each pair related by this transformation.

We introduce the orthonormal basis
\begin{eqnarray}
|0\rangle =\frac{1}{\sqrt{2}}(|\uparrow\downarrow\rangle  -|\downarrow\uparrow \rangle ) \ \ & \ \  |1\rangle =|\uparrow\uparrow \rangle  \nonumber\\
|2\rangle =\frac{1}{\sqrt{2}}(|\uparrow\downarrow\rangle +|\downarrow\uparrow\rangle) \ \ &\ \  |3\rangle =|\downarrow\downarrow\rangle 
\end{eqnarray}
for each rung, and construct a set of corresponding  projection operators which generate the $SU(4)$ algebra (hereafter Greek indices denote 0, 1, 2, 3)
\begin{eqnarray}
X^{\alpha\beta}_i =(|\alpha\rangle\langle\beta|)_i \ \ & \ \ \ \ \ X^{\alpha\beta}_{i}X^{\delta\gamma}_{i}=\delta^{\beta\delta}X^{\alpha\gamma}_{i}    \nonumber\\ 
\sum_{\alpha}X^{\alpha\alpha}_{i}=1 \ \ &\ \ [X^{\alpha\beta}_{i},X^{\gamma\delta}_{j}]= (\delta^{\beta\gamma}X^{\alpha\delta}_{i}-\delta^{\alpha\delta}X^{\gamma\beta}_{j})\delta_{ij} .
\end{eqnarray}
The spin operators ${\bf S}_i$ and ${\bf S}_i$ may be expressed in terms of these projection operators. The ladder Hamiltonian (\ref{ladham}) is thus equivalent to a four-state chain with only nearest-neighbor interactions.    

We introduce the basis of operators 
\begin{eqnarray}
b_{i}=\sum_{a}X^{0a}_{i}X^{a0}_{i+1} &   b_{i}^{\dag}=\sum_{a}X^{a0}_{i}X^{0a}_{i+1}  & B_{i}=\sum_{a,b}X^{ab}_{i}X^{ba}_{i+1} \nonumber\\a_{i}^{\dag}= \sum_{a}(-1)^{a}X_{i}^{a0}X_{i+1}^{(4-a) 0} & a_{i} = \sum_{b}(-1)^{b}X_{i}^{0b}X_{i+1}^{0 (4-b)} &
e_{i}= a_{i}^{\dag} a_{i}=  \sum_{a,b}(-1)^{a+b} X^{ab}_{i}X^{(4-a) (4-b)}_{i+1} \nonumber\\ 
p_{i}^{0}=(1-X^{00}_{i})(1-X^{00}_{i+1}) &  p^{1}_{i}=X_{i}^{00}(1-X^{00}_{i+1})  & p^{2}_{i}=(1-X^{00}_{i})X^{00}_{i+1}\nonumber\\
&  p_{i}^{3}= X^{00}_{i}X^{00}_{i+1}  &  \sum_{\alpha=0}^{3}p_{i}^{\alpha}=I_{i},
\label{algi}
\end{eqnarray}
where Latin indices $a$, $b$ take the values 1, 2, 3 and we have used the plaquette notation $O_{i,i+1}\equiv O_{i}$ for all operators.

This basis provides the full set of operators for isotropic interactions, {\it i.e.} any $SU(2)$-invariant term may be represented by linear combinations of the above generators. The multiplication table of these operators has the form 
\begin{center}
\begin{tabular}{c|cccccccccc}
 &$B_i$ \ & \ $e_i$ \ & \ $b_i$ \ & \ $b_{i}^{\dag}$ \ & \ $a_i$ \ & \ $a_{i}^{\dag}$ \ & \ $p_{i}^{0}$ \ & \ $p_{i}^{1}$ \  & \ $p_{i}^{2}$ \ & \ $p_{i}^{3}$\\
\hline
$B_{i}$ &$p_{i}^{0}$ &$e_{i}$ & 0 & 0 & 0 & $a_{i}^{\dag}$ & $B_{i}$ & 0& 0& 0 \\
$e_{i}$ &$e_{i}$ & $3e_{i}$ & 0 & 0 & 0 & $3a_{i}^{\dag}$ & $e_{i}$ & 0 & 0 & 0 \\
$b_{i}$ & 0 & 0 & 0 & $p_{i}^{1}$ & 0 & 0 & 0 & 0 & $b_{i}$ & 0 \\
$b_{i}^{\dag}$ & 0 & 0 & $p_{i}^{2}$ & 0 & 0 & 0 & 0 & $b_{i}^{\dag}$  & 0 & 0 \\
$a_{i}$ & $a_{i}$ & $3a_{i}$ & 0 & 0 & 0 & $3p_{i}^{3}$ & $a_{i}$ & 0 & 0 & 0 \\
$a_{i}^{\dag}$ & 0 & 0 & 0 & 0 & $e_{i}$ & 0 & 0 & 0 & 0 & $a_{i}^{\dag}$ \\
$p_{i}^{0}$ & $B_{i}$ & $e_{i}$ & 0 & 0 & 0& $a_{i}^{\dag}$ & $p_{i}^{0}$ & 0 & 0 & 0\\
$p_{i}^{1}$ & 0 & 0 & $b_{i}$ & 0 & 0& 0 & 0 & $p_{i}^{1}$ & 0 & 0 \\
$p_{i}^{2}$ & 0 & 0 & 0 & $b_{i}^{\dag}$ & 0& 0 & 0 & 0 & $p_{i}^{2}$ & 0 \\
$p_{i}^{3}$ & 0 & 0 & 0 & 0 & $a_{i}$ & 0 & 0 & 0 & 0 & $p_{i}^{3}$, \\
\end{tabular}
\end{center}
where in a product of two operators of the form $O_{i}^{(1)}O_{i}^{(2)}$, the operator $O_{i}^{(1)}$ is taken from the column on the left whereas the operator $O_{i}^{(2)}$  belongs to the top row.  

In the operator basis defined by Eq.~(\ref{algi}), the Hamiltonian (\ref{ladham}) has the form
\begin{eqnarray}
H=\sum_{i}g_{1} B_{i}+g_{2} e_{i}+g_{3} (b_{i}+b_{i}^{\dag})
+g_{4} (a_{i}+a_{i}^{\dag}) + g_{5} p_{i}^{0} +g_{6} p_{i}^{3}+c,
\label{algham}
\end{eqnarray}
where
\begin{eqnarray}
g_{1} =\frac{J_{L}}{2}+\frac{J_{D}}{2}+\frac{V_{LL}}{8}+\frac{V_{DD}}{8} & &
g_{2} = -\frac{J_{L}}{2}-\frac{J_{D}}{2}+\frac{V_{LL}}{8}+\frac{V_{DD}}{8}\nonumber\\
g_{3} = \frac{J_{L}}{2}-\frac{J_{D}}{2}+\frac{V_{LL}}{8}-\frac{V_{DD}}{8} & & 
g_{4} = \frac{J_{L}}{2}-\frac{J_{D}}{2}-\frac{V_{LL}}{8}+\frac{V_{DD}}{8}\nonumber\\
g_{5} = \frac{V_{RR}}{4}+\frac{J_{R}}{2} & & 
g_{6} = -\frac{J_{R}}{2}+\frac{V_{LL}}{4}+\frac{V_{DD}}{4}+\frac{3 V_{RR}}{4}\nonumber\\
& & c= -\frac{J_{R}}{4}-\frac{V_{LL}}{16}-\frac{V_{DD}}{16}-\frac{3 V_{RR}}{16}
\end{eqnarray}

\subsection{Operator algebra}
 
The operators defined in (\ref{algi}) generate a closed algebra whose commutation relations
for operators defined on the same plaquette are given in the multiplication table, while for operators on neighboring plaquettes one obtains
\begin{center}
\begin{eqnarray}
e_{i}e_{i+1}e_{i}=e_{i}p_{i+1}^{0} \ & \ e_{i+1}e_{i}e_{i+1}=p_{i}^{0}e_{i+1} \nonumber\\
B_{i}B_{i+1}B_{i}=B_{i+1}B_{i}B_{i+1} \ & \ e_{i}B_{i+1}e_{i}=e_{i}p_{i+1}^{0}\nonumber\\
B_{i}e_{i+1}=B_{i+1}e_{i}e_{i+1} \ &  \  e_{i}B_{i+1}=e_{i}e_{i+1}B_{i}\nonumber\\
B_{i}e_{i+1}B_{i}=B_{i+1}e_{i}B_{i+1} 
\label{eBalg}
\end{eqnarray}
for operators $e_{i}$, called monoids and $B_{i}$, called braid operators, and
\begin{eqnarray}
b_{i}^{\dag}a_{i+1}^{\dag}=b_{i+1}a_{i}^{\dag} \ & \  b_{i}^{\dag}a_{i+1}^{\dag}a_{i}=b_{i+1}e_{i} \nonumber\\
b_{i}^{\dag}b_{i+1}^{\dag}=a_{i+1}a_{i}^{\dag} \ & \  b_{i}^{\dag}b_{i+1}^{\dag}a_{i}=a_{i+1}e_{i} \nonumber\\  
b_{i}^{\dag}e_{i+1}=b_{i+1}a_{i}^{\dag}a_{i+1} \ & \  b_{i}^{\dag}e_{i+1}b_{i}=b_{i+1}e_{i}b_{i+1}^{\dag} \nonumber\\
& b_{i}^{\dag}B_{i+1}b_{i}=b_{i+1}B_{i}b_{i+1}^{\dag}  
\label{alg1}
\end{eqnarray}
\begin{eqnarray}
b_{i}b_{i+1}a_{i}^{\dag}=a_{i+1}^{\dag}p_{i}^{3} \ & \  b_{i}a_{i+1}e_{i}=b_{i+1}^{\dag}a_{i}  \nonumber\\
b_{i}a_{i+1}a_{i}^{\dag}=p_{i}^{1}b_{i+1}^{\dag} \ & \  b_{i}a_{i+1}B_{i}= b_{i+1}^{\dag}a_{i}B_{i+1} \nonumber\\
& b_{i}b_{i+1}B_{i}= B_{i+1}b_{i}b_{i+1}  
\end{eqnarray}
\begin{eqnarray}
a_{i}^{\dag}b_{i+1}b_{i}=e_{i}a_{i+1}^{\dag} \ & \  a_{i}^{\dag}a_{i+1}b_{i}=e_{i}b_{i+1}^{\dag}  \nonumber\\
a_{i}^{\dag}b_{i+1}=e_{i}a_{i+1}^{\dag}b_{i} \ & \  a_{i}^{\dag}a_{i+1}=e_{i}b_{i+1}^{\dag}b_{i}^{\dag}
\end{eqnarray}
\begin{eqnarray}
a_{i}a_{i+1}^{\dag}=b_{i+1}b_{i} \ & \  a_{i}a_{i+1}^{\dag}b_{i}^{\dag}=b_{i+1}p_{i}^{1} \nonumber\\
 a_{i}b_{i+1}^{\dag}=a_{i+1}b_{i} \ & \  a_{i}b_{i+1}^{\dag}b_{i}^{\dag}=a_{i+1}p_{i}^{1} \nonumber\\  
a_{i}e_{i+1}=a_{i}B_{i+1}B_{i}=b_{i+1}b_{i}a_{i+1} \ & \  a_{i}B_{i+1}e_{i}=a_{i}p_{i+1}^{0} \nonumber\\
a_{i}e_{i+1}e_{i}=a_{i}B_{i+1}e_{i} = a_{i}p_{i+1}^{0} \  & \  a_{i}e_{i+1}B_{i}=a_{i}B_{i+1} \nonumber\\
&  a_{i}e_{i+1}a_{i}^{\dag}=a_{i}B_{i+1}a_{i}^{\dag} = p_{i}^{3}p_{i+1}^{1}     
\end{eqnarray}
\begin{eqnarray}
e_{i}a_{i+1}^{\dag}=a_{i}^{\dag}b_{i+1}b_{i}=B_{i+1}B_{i}a_{i+1}^{\dag}  &   e_{i}e_{i+1}a_{i}^{\dag}=a_{i}^{\dag}p_{i+1}^{1}\nonumber\\
e_{i}a_{i+1}^{\dag}b_{i}^{\dag}=a_{i}^{\dag}b_{i+1} \ & \  e_{i}b_{i+1}^{\dag}=a_{i}^{\dag}a_{i+1}b_{i}\nonumber\\   
e_{i}b_{i+1}^{\dag}b_{i}^{\dag}=a_{i}^{\dag}a_{i+1} \ & \   e_{i}B_{i+1}a_{i}^{\dag}=a_{i}^{\dag}p_{i+1}^{1} 
\label{DTL}
\end{eqnarray}
\begin{eqnarray}
B_{i}a_{i+1}^{\dag}=B_{i+1}e_{i}a_{i+1}^{\dag} \ & \ B_{i}a_{i+1}^{\dag}b_{i}^{\dag}=B_{i+1}a_{i}^{\dag}b_{i+1} \nonumber\\
B_{i}e_{i+1}a_{i}^{\dag}=B_{i+1}a_{i}^{\dag} \ & \  B_{i}B_{i+1}a_{i}^{\dag}=e_{i+1}a_{i}^{\dag}  \nonumber\\
B_{i}b_{i+1}^{\dag}b_{i}^{\dag}=b_{i+1}^{\dag}b_{i}^{\dag}B_{i+1}. 
\label{algf}
\end{eqnarray}
\end{center}
between operators of different kind. For $|i-j|>1$ all operators commute.

This algebra provides a representation of the dilute two-color braid-monoid algebra introduced and investigated in \cite{Grimm1}-\cite{Grimm5}, or, more explicitly, to its undeformed limit. It emerges in the investigation of dilute $A$-$D$-$E$ models, and may be considered as a special case of the two-color generalization of the braid-monoid algebra \cite{WDA}, where the representation of one of the colors is trivial. This algebra possesses a graphical interpretation in terms of strands or strings, such that the relations of the algebra are equivalent to the deformation of the diagrams.

The operators $e_{i}$ satisfy the relations of the Temperley-Lieb algebra \cite{TL} (see the first line in Eq.~(\ref{eBalg})). The braid generators $B_{i}$, together with $e_{i}$, satisfy the braid-monoid (or Brauer \cite{Brauer}) algebra (\ref{eBalg}), a subalgebra of the dilute two-color braid-monoid algebra. Later we will also introduce full braid and full monoid generators which satisfy the $O(4)$-related Brauer algebra relations. 
All of these algebraic relations permit the construction of different Baxterization patterns.  

\subsection{Baxterization}

The Yang-Baxter equation plays a central role in the study of integrable models. By starting from the braid-group representation one may construct its solutions, which depend on the spectral parameter, and demand that in the limit, as the spectral parameter is taken to be zero, the given braid-group representation is recovered. The process of seeking this solution is called Baxterization. It was introduced by Jones in Ref.~\cite{Jones}, where it was found that the  Bir\-man-Wen\-zl-Mu\-ra\-ka\-mi algebra provides the solution of Yang-Baxter equation. In the present considerations, the algebraic relations (\ref{alg1}-\ref{algf})  permit the construction of different Baxterization patterns.

The solution of the Yang-Baxter equation
\begin{eqnarray}
\check{R}_{i}(u)\check{R}_{i+1}(u+v)\check{R}_{i}(v)=\check{R}_{i+1}(v)\check{R}_{i}(u+v)\check{R}_{i+1}(u),
\end{eqnarray}
where $u$ is real, is sought in the form
\begin{eqnarray}
 \check{R}_{i}(u)=a(u)(a_{i}+a_{i}^{\dag})+ b(u)(b_{i}+b_{i}^{\dag}) + c(u)B_{i}+d(u)e_{i} \nonumber\\ 
+e(u)p_{i}^{0}+ f(u)(p_{i}^{1}+p_{i}^{2})+g(u)p_{i}^{3}.
\end{eqnarray}
Following the standard analysis we require the initial condition, $\check{R}_{i}(0)= I_{i}$, and the unitarity condition, $\check{R}_{i}(u)\check{R}_{i}(-u)=k(u)k(-u)I_{i}$, for some function $k(u)$ which we demand to be  either rational, trigonometric or hyperbolic (in this work).
Following the  algebraic Bethe ansatz technique (see e.g. Ref.~\cite{Fad} for a review), the corresponding soluble Hamiltonian is constructed as 
\begin{eqnarray}
H = \sum_{i}\frac{d \check{R}_{i}(u)}{d u}|_{u=0}.
\end{eqnarray}
This equation amounts to a strong condition on the general Hamitonian (\ref{algham}) and thus restricts the values of coupling parameters, for which the model is integrable, to a few points and lines in parameter space.

\section{Exact solutions}

We study different particular cases by taking certain operator combinations. The set of ``basic'' generators $\{ B_{i}, e_{i}, a_{i},a_{i}^{\dag}, b_{i}, b_{i}^{\dag}\}$ form a basis for all interaction terms in (\ref{ladham}) except for rung and rung-rung interactions. The hermiticity of the Hamiltonian requires that operators ($a_{i}$,  $a^{\dag}_{i}$) and  ($b_{i}$, $b_{i}^{\dag}$) enter only as the sums $a_{i}+a_{i}^{\dag}$ and $b_{i}+b_{i}^{\dag}$. Simple combinatorial arguments give the total possible number of exactly soluble models. Four combinations may be constructed from single generators ({\it i.e.} $B_{i}$, $e_{i}$, $a_{i}+a^{\dag}_{i}$ and $b_{i}+ b_{i}^{\dag}$), there are six pair combinations, four triple combinations and one combination constructed from all four generators. All of these cases must be considered together with the projectors $p_{i}^{\alpha}$. The combination $a_{i}+a_{i}^{\dag}$ enters only together with the monoid $e_{i}$, as otherwise the unitarity condition for the $\check{R}$ matrix is not satisfied. This excludes one single, two pairs and one triple combination. Thus there are in total 11 possible combinations which may potentially yield exactly soluble models, but we will show that not all of these are relevant to the spin-1/2 ladder or possess a Baxterization Ansatz.
For example, two triple combinations, $\{ B_{i}, b_{i}+b_{i}^{\dag}, e_{i}\}$ and  $\{ B_{i}, a_{i}+a_{i}^{\dag}, e_{i}\}$, appear not to be integrable. In these cases the integrability test \cite{GM} for the local Hamiltonian $h_{i,i+1}$, 
\begin{eqnarray}
\sum_{i}[h_{i,i+1}+h_{i+1,i+2},[h_{i,i+1},h_{i+1,i+2}]]=0,
\end{eqnarray}
is not satisfied, whereas for all the integrable cases investigated in the present work it is satisfied. 

For certain cases we will present particular discrete transformations which increases the number of soluble Hamitonians.

\subsection{ The combination $\{ b_{i}, b_{i}^{\dag}, p_{i}^{\alpha}\}$}
This combination gives the solution to the Yang-Baxter equation for which
the corresponding $R$-matrix has the form
\begin{equation}
\check{R}_{i}(u)= I_{i}+u(b_{i}+b_{i}^{\dag}+p_{i}^{0}+p_{i}^{3}).
\label{R1/2}
\end{equation}
This solution has exactly the form of the spin-1/2 Heisenberg model. Indeed, the composite operator $\Pi_{i}= b_{i}+b_{i}^{\dag}+p_{i}^{0}+p_{i}^{3}$ satisfies the braid algebra relation $\Pi_{i}\Pi_{i+1}\Pi_{i}=\Pi_{i+1}\Pi_{i}\Pi_{i+1}$ and its square gives the identity operator. The Hamiltonian which follows from this $R$-matrix has exactly the same eigenspectrum as the spin-1/2 Heisenberg model, but the degeneracy of this spectrum is different \cite{Alcaraz}. The corresponding Hamiltonian 
$H_{1/2}= \sum_{i}\Pi_{i}$
commutes with the generators $\sum_{i}X^{ab}_{i}$ for any $a,b=1,2,3$. These generators form an $su(3)$ subalgebra within the $su(4)$ algebra generated by the $X$-operators. $H_{1/2}$ commutes with the operator $\sum_{i}X^{00}_{i}$, which represents the total number of rung singlets, and any multiple of this term may be added to the Hamiltonian without affecting the integrability. It corresponds to a coupling of the effective spin-1/2 Heisenberg model to an effective ``magnetic field" $h$. It is well known  that the spin-1/2 chain in a magnetic field develops an incommensurate critical phase for $|h|\leq2$ and has a  massive phase for $|h|>2$. Returning to the original spin variables, we have
\begin{eqnarray}
H_{1/2}  =  
  \sum_{i}  {\bf S}_{i}\cdot{\bf S}_{i+1} + {\bf T}_{i}\cdot{\bf T}_{i+1} 
-{\bf S}_{i}\cdot{\bf T}_{i+1} - {\bf T}_{i}\cdot{\bf S}_{i+1} \nonumber\\
+4[({\bf S}_{i}\cdot{\bf S}_{i+1})({\bf T}_{i}\cdot{\bf T}_{i+1})
+({\bf S}_{i}\cdot{\bf T}_{i})({\bf S}_{i+1}\cdot{\bf T}_{i+1})  \nonumber\\
-({\bf S}_{i}\cdot{\bf T}_{i+1})({\bf S}_{i+1}\cdot{\bf T}_{i})]+
 (J_{R}+2) ({\bf S}_{i}\cdot{\bf T}_{i}),
\label{mag}
\end{eqnarray}
where $J_{R}=0$ correspond to the Hamiltonian derived from the $\check{R}$-matrix (\ref{R1/2}).
The Hamiltonian commutes with $\sum_{i}X_{i}^{00}$, and remains therefore integrable for arbitrary $J_{R}$. For $-4\leq J_{R}\leq 4$ the model is in the critical, incommensurate phase, while for $J_{R}>4$ it has a gapped rung-singlet phase, and in the region  $J_{R}<-4$ it has a rung-triplet ground state. It coincides with the model studied in Ref.~\cite{Alcaraz} and more recently in Ref.~\cite{AFW}.  Although the $\check{R}$-matrix of Ref.~\cite{AFW} is not the same as in (\ref{R1/2}), we believe that they are related by an appropriate transformation.   

In the critical region the model is described by a conformal field theory with  central charge $c=1$. However,  because of the degeneracy this conformal field theory possesses additional zero modes, and therefore the conformal dimensions 
appearing in this model have the Coulomb-gas dimensions of the $c=1$ theory, but with different degeneracies.

\subsection{The combination $\{ b_{i}, b_{i}^{\dag}, B_{i},  p^{\alpha}_{i} \}$}
This combination leads to the $\check{R}$-matrix
\begin{eqnarray}
\check{R}_{i}(u)= I_{i}+u(\pm(b_{i}+b_{i}^{\dag})+ B_{i}+ p_{i}^{3}),
\end{eqnarray}
a solution which corresponds simply to local Hamiltonian operators proportional to the permutation operators (see Sec. 3.5 below) $P^{||}_{i}$ (for positive sign) and $P^{\times}_{i}$ (for negative sign). These Hamiltonians are invariant with respect to the full $SU(4)$ group, and describe the soluble ``spin-orbital"  model \cite{KK}
\begin{eqnarray}
H_{P^{||}}=\sum_{i} {\bf S}_{i}\cdot{\bf S}_{i+1} + {\bf T}_{i}\cdot{\bf T}_{i+1} 
+4({\bf S}_{i}\cdot{\bf S}_{i+1})({\bf T}_{i}\cdot{\bf T}_{i+1}). 
\label{||}
\end{eqnarray}
The model corresponding to the $P^{\times}_{i}$ can be obtained from the (\ref{||}) by applying the transformation (\ref{inter}).
Both Hamiltonians commute with $\sum_{i}X_{i}^{00}$, and therefore the term $J_{R}\sum_{i}{\bf S}_{i}\cdot{\bf T}_{i}$  may be added to these Hamiltonians without affecting the property of integrability.  The model (\ref{||}) was studied in Ref.~\cite{Wang} for certain values of the rung-rung interaction parameter $V_{RR}$ added to the Hamiltonian. 
For some regions in the parameter space of $J_{R}$ and $V_{RR}$, these models demonstrate a variety of gapless behavior. In particular, the model (\ref{||}) in the continuum limit corresponds to the $SU(4)$ Wess-Zumino-Witten (WZW) model at level 1 of the Kac-Moody algebra \cite{AGLN}, \cite{IQA}. The central charge is equal to 3.

\subsection{The combination $ \{ b_{i}, b_{i}^{\dag}, e_{i},  p^{\alpha}_{i} \}$}
This yields the  hyperbolic solution
\begin{eqnarray}
\check{R}_{i}(u)=\frac{\sinh(\lambda -u)}{\sinh(\lambda)}(p_{i}^{0}+p_{i}^{3})+
(p_{i}^{1}+p_{i}^{2})+\frac{\sinh(u)}{\sinh(\lambda)}(e_{i}+\sigma (b_{i}+b_{i}^{\dag})),
\end{eqnarray}
with $\cosh(\lambda)=3/2$ and $\sigma =\pm 1$.

The corresponding Hamiltonian is
\begin{eqnarray}
H=\sum_{i} 
-{\bf S}_{i}\cdot{\bf T}_{i+1} - {\bf T}_{i}\cdot{\bf S}_{i+1} +4[({\bf S}_{i}\cdot{\bf S}_{i+1})({\bf T}_{i}\cdot{\bf T}_{i+1})\nonumber\\- ({\bf S}_{i}\cdot{\bf T}_{i})({\bf S}_{i+1}\cdot{\bf T}_{i+1})] + J_{R} ({\bf S}_{i}\cdot{\bf T}_{i}) 
\end{eqnarray}
for  $\sigma =+1$ and $J_{R}=-1$, and the case of  $\sigma =-1$ is obtained by the transformation (\ref{inter}).
We note that these Hamiltonians commute with the rung singlet number operator $\sum_{i}X_{i}^{00}$, and that the model is therefore integrable for any $J_{R}$.

\subsection{The combination  $ \{ b_{i}, b_{i}^{\dag}, e_{i}, a_{i},a_{i}^{\dag},  p^{\alpha}_{i} \}$}
This combination also permits Baxterization. The solution is given by the Izergin-Korepin model, which is related to the dilute $A_{2}^{(2)}$ model \cite{IzKor},\cite{War}.
\begin{eqnarray}
\check{R}_{i}(u)= \frac{\sin(2\lambda -u)\sin(3\lambda -u)}{\sin(2\lambda)\sin(3\lambda)}p_{i}^{0}+\frac{\sin(3\lambda -u)}{\sin(3\lambda)}(p_{i}^{1}+p_{i}^{2})+
(1+\frac{\sin(u)\sin(3\lambda -u)}{\sin(2\lambda)\sin(3\lambda)}) p_{i}^{3}\nonumber\\
+ \frac{\sin(u)\sin(3\lambda -u)}{\sin(2\lambda)\sin(3\lambda)} (b_{i}^{\dag}+b_{i})
- \frac{\sin(u)\sin(\lambda -u)}{\sin(2\lambda)\sin(3\lambda)}e_{i}+\frac{\sin(u)}{\sin(3\lambda)}(a_{i}^{\dag}+a_{i})
\end{eqnarray}
 However, the value of the Temperley-Lieb factor (or fugacity in corresponding statistical mechanics models) $n$ in the relation $e_{i}^{2}= n e_{i}$ is restricted here by the condition $n=-2\cos(4\lambda)$. In our case $n=3$ (see multiplication table) and therefore $\lambda$ is necessary complex. The analytic continuation of the weights into the domain $|n|> 2$ by the change $\lambda\rightarrow \pi/4+i\tilde{\lambda}$ and $u\rightarrow i\tilde{u}$ makes these complex,  and the corresponding Hamiltonian then has complex coefficients which make this operator combination not relevant for the spin-1/2 ladder model. 

\subsection{The full set of operators}
The full set of operators provides solutions related to the Brauer algebra \cite{Brauer}, or undeformed braid-monoid algebra \cite{WDA}. In order to present this solution we introduce full braid and monoid operators, whose relations with the generators (\ref{algi}) of the  dilute algebra are given by
\begin{eqnarray}
E_{i}^{(-)} = e_{i} + a_{i}+a_{i}^{\dag}+p_{i}^{3} \ & \ E_{i}^{(+)}= e_{i} - a_{i}-a_{i}^{\dag}+p_{i}^{3} \nonumber\\
P_{i}^{||}=B_{i}+b_{i}+b_{i}^{\dag}+p_{i}^{3} \ & \ P_{i}^{\times}=B_{i}-b_{i}-b_{i}^{\dag}+p_{i}^{3} 
\label{full}
\end{eqnarray}
where $P^{||}_{i,i+1}$ and $P^{\times }_{i,i+1}$  are respectively the permutation operators corresponding to leg-leg and diagonal-diagonal bonds.

The operators $E_{i,i+1}^{(-)}$, $E_{i,i+1}^{(+)}$, $P^{||}_{i,i+1}$ and  $P^{\times }_{i,i+1}$ are generators of the algebra
\begin{eqnarray}
 \nonumber\\
E_{i}^2 =4E_{i}   \ & \ \  E_{i}E_{i+1}E_{i}=  E_{i}  \ & \  E_{i+1}E_{i}E_{i+1}= E_{i+1} \nonumber\\ 
P_{i}^2 =1   \ & \ \  P_{i}P_{i+1}P_{i}=  P_{i+1}P_{i}P_{i+1}  &    P_{i}E_{i}=E_{i}P_{i}=E_{i}   \nonumber\\
P_{i}E_{i+1}E_{i}=  P_{i+1}E_{i} \ \ & \ \  E_{i+1}E_{i}P_{i+1}=  E_{i+1}P_{i} \ & \ P_{i}E_{i+1}P_{i}=  P_{i+1}E_{i}P_{i+1}\nonumber\\
& \mbox{and} & \nonumber\\
 P_{i}P_{j}=  P_{j}P_{i} \ & \  E_{i}E_{j}=  E_{j}E_{i} \ & \  P_{i}E_{j}=E_{j}P_{i}  \nonumber\\
& \mbox{for} \ |i-j|>1. &
\label{Balg}
\end{eqnarray}
We have again used the plaquette notation $P_{i,i+1}\equiv P_{i}$ and $E_{i,i+1}\equiv E_{i}$, where $P_{i}$ denotes either $P^{||}_{i}$ or $P^{\times }_{i}$ 
and $E_{i}$ denotes either $E_{i}^{(-)}$ or $E_{i}^{(+)}$. Note that there exist a number of relations between the operators which follow from the above. The operators $P_{i}$  satisfy the braid algebra relations, while the operators $E_{i}$ form a Temperley-Lieb algebra \cite{TL}. Together these operators form a Brauer \cite{Brauer} (or braid monoid \cite{WDA}) algebra. The operators $E_{i}^{(\pm)}$ are projectors on the plaquette singlet states (below).

This algebra allows Baxterization \cite{WDA},\cite{Cheng}, the solution for the case under consideration being given by the rational $\check{R}$ matrix
\begin{eqnarray}
\check{R}_{i}(u)= I_{i}+ u P_{i} - \frac{u}{u+1}E_{i},
\label{BR}
\end{eqnarray}
which satisfies the initial and unitarity conditions with $k(u)= 1+u$.
The spin Hamiltonian which corresponds to this R matrix is
\begin{eqnarray} 
H=\sum_{i}(P_{i}-E_{i}).
\end{eqnarray}
We thus obtain four different soluble Hamiltonians which correspond to the four combinations  $\{ P^{||}_{i},E_{i}^{(-)}\}$, $\{ P^{||}_{i},E_{i}^{(+)}\}$, $\{ P^{\times}_{i},E_{i}^{(-)}\}$, and $\{ P^{\times}_{i},E_{i}^{(+)}\}$.
Two of these are trivially soluble, because  $\sum_{i} (P^{||}_{i}-E^{(+)}_{i})$ is the Hamiltonian of two decoupled chains and $\sum_{i}(P^{\times}_{i}-E_{i}^{(-)})$ is the same pair of chains intertwined by the transformation (\ref{inter}). Of the remaining two, one is given by the combination $\{ P^{||}_{i},E_{i}^{(-)}\}$, which in terms of spin operators is  
\begin{eqnarray}
H_B = \sum_{i}{\bf S}_{i}\cdot{\bf S}_{i+1} + {\bf T}_{i}\cdot{\bf T}_{i+1} + {\bf S}_{i}\cdot{\bf T}_{i+1} + {\bf T}_{i}\cdot{\bf S}_{i+1} \nonumber\\ 
+4[({\bf S}_{i}\cdot{\bf S}_{i+1})({\bf T}_{i}\cdot{\bf T}_{i+1}) - ({\bf S}_{i}\cdot{\bf T}_{i+1})({\bf S}_{i+1}\cdot{\bf T}_{i})]
\label{HB}
\end{eqnarray}  
The last, $\{ P^{\times}_{i},E_{i}^{(+)}\}$, is obtained from the previous Hamiltonian by the same intertwining transformation (\ref{inter}).

The Brauer algebra (\ref{Balg}) is related to the representations of the group $D_{2}=O(4)$, and therefore  the corresponding Bethe Ansatz is also related to this algebra. In fact the corresponding operator $\cal{L}$ is the product of two $\cal{L}$-operators for 6-vertex models. The eigenvalues of the algebraic Bethe Ansatz are given by the product of two sets of eigenvalues corresponding to 6-vertex models \cite{Martins}, and finally one obtains two decoupled Bethe Ansatz equations which correspond to the $D_2=SU(2)\times SU(2)$ algebra. The corresponding energy spectrum is the sum of the eigenvalues for each $SU(2)$ component. Thus the Hamiltonian (\ref{HB}) is equivalent to the Hamiltonian of two decoupled chains. This model is critical (no spin gap) with a central charge $c= 2$.

The two projectors $E_{i}^{(-)}$ and $E_{i}^{(+)}$ are related by the $X_i$-operator transformation
\begin{eqnarray}
X^{0a}_i\rightarrow - i X_{i}^{0a} \ & \ X^{a0}_i\rightarrow i X_{i}^{a0} 
\end{eqnarray}
for all sites $i=1,..,N$. This is a unitary transformation generated by the operator
\begin{eqnarray}
U(\pi/2)= \exp[-i\frac{\pi}{2}\sum_{i=1}^N(X^{00}_i )]. 
\label{Z4}
\end{eqnarray}
It leaves $P_{i}$ invariant and transforms the Hamiltonian of two decoupled chains into $H_{B}$. Moreover, because of the relation $[U(\pi/2)]^4 =1$, this transformation is one of the generators of the $Z_{4}$ group. 

The projectors may be decomposed into the product of two operators, namely
\begin{eqnarray}
E_{i}^{(-)}= A_{i}^{\dag}A_{i} \ & \  E_{i}^{(+)}=B_{i}^{\dag} B_{i} ,
\end{eqnarray}
where
\begin{eqnarray}
A_{i}^{\dag}&=&X_{i}^{00}X_{i+1}^{00}+X_{i}^{20}X_{i+1}^{20}- X_{i}^{10}X_{i+1}^{30}- X_{i}^{30}X_{i+1}^{10}\nonumber\\
B_{i}^{\dag}&=&X_{i}^{00}X_{i+1}^{00}-X_{i}^{20}X_{i+1}^{20}+ X_{i}^{10}X_{i+1}^{30}+ X_{i}^{30}X_{i+1}^{10}.
\label{ABoper}
\end{eqnarray}
These operators obey the relations
\begin{eqnarray}
A_{i}^{2}=A_{i} \ & \ B_{i}^{2}=B_{i}  \nonumber\\
 A_{i}A_{i}^{\dag} = 4 X^{00}_{i}X^{00}_{i+1} \ & \  B_{i} B_{i}^{\dag} = 4 X^{00}_{i}X^{00}_{i+1}. 
\end{eqnarray}
The transformation (\ref{Z4}) yields  $A_{i}\leftrightarrow B_{i}$ and $E_{i}^{(-)}\leftrightarrow E_{i}^{(+)}$.
The unnormalized projectors may be expressed as
\begin{eqnarray}
E_{i}^{(-)} = |\psi^{(-)}_{i}\rangle\langle\psi^{(-)}_{i}| \ & \ E_{i}^{(+)}  = |\psi^{(+)}_{i}\rangle\langle\psi^{(+)}_{i}| ,
\end{eqnarray}
where
\begin{eqnarray}
|\psi^{(-)}_{i}\rangle= A^{\dag}_{i} |0\rangle \ \ & \ \ |\psi^{(+)}_{i}\rangle = B^{\dag}_{i} |0\rangle ,
\label{plaqstat}
\end{eqnarray}
and we have defined the vacuum state as a direct product of singlet states for all rungs.
The states
$|\psi^{(-)}_{i}\rangle$ and $|\psi^{(+)}_{i}\rangle$  may be expressed in terms of local singlet states $|s_{ij}\rangle$ ($i,j=1,2,3,4$) on the bonds of each plaquette. Using the $X$-operator notation and denoting the plaquette vertices in clockwise order, one may write 
\begin{eqnarray}
|\psi^{(-)}_{i}\rangle=-2(|s_{12}\rangle |s_{34}\rangle -|s_{23}\rangle |s_{41}\rangle )  \ & \
|\psi^{(+)}_{i}\rangle=-2|s_{23}\rangle |s_{41}\rangle .
\end{eqnarray}
These functions yield a plaquette-singlet state, {\it i.e.}  the state with the total spin per plaquette $({\bf S}_{1}+{\bf S}_{2}+{\bf S}_{3}+{\bf S}_{4})^{2}$ equal to zero. There are exactly two possible plaquette-singlet states which correspond respectively to $E_{i}^{(-)}$ and $E_{i}^{(+)}$. The $Z_{4}$ transformation (\ref{Z4}) maps one state into the other.

In terms of the original spin variables this transformation is given by \cite{dual}
\begin{eqnarray}
{\bf\tilde{S}}_{i}=\frac{1}{2}({\bf S}_{i}+{\bf T}_{i})- {\bf S}_{i}\times {\bf T}_{i} \ \ & \ \ 
{\bf\tilde{T}}_{i}=\frac{1}{2}({\bf S}_{i}+{\bf T}_{i})+ {\bf S}_{i}\times {\bf T}_{i}
\label{dual}
\end{eqnarray}
The transformation is canonical for the spin-1/2 $\otimes$ spin-1/2 representation of $SU(2)\times SU(2)$ group and the values of the Casimir operators are the same, ${\bf\tilde{S}}_{i}^{2}={\bf\tilde{T}}_{i}^{2}=3/4$. 
In terms of the variables ${\bf\tilde{S}}_{i}, {\bf\tilde{T}_{i}}$ the Hamiltonian $H_{B}$ (\ref{HB}) indeed has the form of two decoupled spin chains.

\subsection{The full monoid generators $E_{i}^{(-)}$ and $E_{i}^{(+)}$}
These combinations allow Baxterization according to the Ansatz for the Temperley-Lieb algebra,
\begin{eqnarray}
\check{R}_{i}(u)= \frac{\sinh(\lambda -u)}{\sinh{\lambda}}I_{i}+\frac{\sinh(u)}{\sinh{\lambda}}E_{i},
\end{eqnarray}
where $\cosh(\lambda)=2$. Here $E_{i}$ denotes either $E_{i}^{(-)}$ or $E_{i}^{(+)}$ defined in  Eq.~(\ref{full}).
The corresponding Hamiltonians are
\begin{eqnarray}
H = -\sum_{i}\{ {\bf S}_{i}\cdot{\bf T}_{i+1} + {\bf T}_{i}\cdot{\bf S}_{i+1} 
-4({\bf S}_{i}\cdot{\bf T}_{i+1})({\bf S}_{i+1}\cdot{\bf T}_{i})\}
\label{-}
\end{eqnarray}
for the full Temperley-Lieb generator $E_{i}^{(-)}$, and 
\begin{eqnarray}
H = -\sum_{i} \{ {\bf S}_{i}\cdot{\bf S}_{i+1} + {\bf T}_{i}\cdot{\bf T}_{i+1}  
-4({\bf S}_{i}\cdot{\bf S}_{i+1})({\bf T}_{i}\cdot{\bf T}_{i+1})\}
\label{+}
\end{eqnarray}
for generator  $E_{i}^{(+)}$. We note that these Hamiltonian operators are related by a $Z_{4}$ transformation. In this particular case two Hamiltonians (\ref{-}) and (\ref{+}) may also be related by the transformation (\ref{inter}). The spectrum of both Temperley-Lieb soluble models is gapped and their ground-state energy and gap may in principle be calculated by a mapping onto the Potts model \cite{Batch}.

\subsection{Models related to the bilinear-biquadratic spin-1 chain}

In this subsection we project the space of states onto a three-state subspace by applying the projection operator $p_{i}^{0}$ on every site. This gives three soluble models which are related to the generalized spin-1 chain. The role of unit operator $I_{i}$ is played by the operator $p_{i}^{0}$. The realization of these models in the spin-1/2 ladder is discussed below.

\subsubsection{The combination $\{ B_{i}, e_{i}, p_{i}^{0}\}$} 
This combination of the braids and monoids of the dilute algebra (\ref{algi}) form the Brauer algebra (\ref{Balg}) in a restricted 3-state subspace with Temperley-Lieb (fugacity) factor $n =3$ (in the relation $e_{i}^{2}=n e_{i}$), and leads to the solution
\begin{eqnarray}
\check{R}_{i}(u)= I_{i}^{(3)}+ u B_{i} - \frac{u}{u+1/2}e_{i}.
\end{eqnarray}
This $\check{R}$-matrix solution corresponds to the Takhtajan-Babudjian spin-1 integrable model \cite{BT}.
The corresponding Hamiltonian is
\begin{eqnarray}
H_{TB}= \sum_{i} \frac{3}{2}( {\bf S}_{i}\cdot{\bf S}_{i+1} + {\bf T}_{i}\cdot{\bf T}_{i+1} 
+{\bf S}_{i}\cdot{\bf T}_{i+1} + {\bf T}_{i}\cdot{\bf S}_{i+1}) \nonumber\\
-2[({\bf S}_{i}\cdot{\bf S}_{i+1})({\bf T}_{i}\cdot{\bf T}_{i+1})
+({\bf S}_{i}\cdot{\bf T}_{i+1})({\bf S}_{i+1}\cdot{\bf T}_{i})].  \nonumber\\
\end{eqnarray}
This model  is described by a $c=3/2$, $SU(2)_{k=2}$ WZW model in the continuum limit.

\subsubsection{The dilute braid generator $B_{i}$} 
This generator is a permutation operator in the 3-state subspace and leads to the $SU(3)$-invariant Uimin-Lai-Sutherland \cite{ULS} model,    
\begin{eqnarray}
\check{R}_{i}(u)= I_{i}^{(3)}+ u B_{i}. 
\end{eqnarray}
The spin Hamiltonian is 
\begin{eqnarray}
H_{ULS}= \sum_{i}\frac{1}{2}[{\bf S}_{i}\cdot{\bf S}_{i+1} + {\bf T}_{i}\cdot{\bf T}_{i+1} 
+{\bf S}_{i}\cdot{\bf T}_{i+1} + {\bf T}_{i}\cdot{\bf S}_{i+1}] \nonumber\\
+2[({\bf S}_{i}\cdot{\bf S}_{i+1})({\bf T}_{i}\cdot{\bf T}_{i+1})
+({\bf S}_{i}\cdot{\bf T}_{i+1})({\bf S}_{i+1}\cdot{\bf T}_{i})].  
\end{eqnarray}
In the continuum limit it is described by the $SU(3)_{k=1}$ WZW model \cite{Aff}.

\subsubsection{The dilute monoid operator $e_{i}$}
This operator generates the Temperley-Lieb algebra in the 3-state restricted subspace (see relations (\ref{algi}), (\ref{DTL})), and leads to the solution given by the usual Baxterization Ansatz for the Temperley-Lieb algebra,
\begin{eqnarray}
\check{R}_{i}(u)= \frac{\sinh(\lambda -u)}{\sinh{\lambda}}I_{i}^{(3)}+ \frac{\sinh(u)}{\sinh{\lambda}} e_{i} ,
\end{eqnarray}
where $\cosh(\lambda)=3/2$. 
The corresponding Hamiltonian 
\begin{eqnarray}
H=\frac{1}{2}\sum_{i}-{\bf S}_{i}\cdot{\bf S}_{i+1} - {\bf T}_{i}\cdot{\bf T}_{i+1} 
-{\bf S}_{i}\cdot{\bf T}_{i+1} - {\bf T}_{i}\cdot{\bf S}_{i+1} \nonumber\\
+4[({\bf S}_{i}\cdot{\bf S}_{i+1})({\bf T}_{i}\cdot{\bf T}_{i+1})
+({\bf S}_{i}\cdot{\bf T}_{i+1})({\bf S}_{i+1}\cdot{\bf T}_{i})]\nonumber\\
\end{eqnarray}
is related to the spin-1 chain model studied in Ref.~\cite{Batch}.

\subsubsection{Generalization}
The three solutions above correspond to soluble, spin-1, bilinear-biquadratic chains. Their relation to the ladder (with four state per rung) is the following: on rung $i$ the effective spin-1 variable is formed by the generators $X_{i}^{a0}$, $(a=1,2,3)$. If there is a finite region on the ladder formed by
rung triplets bounded by rung singlets, $X^{00}_{i-1}X^{a0}_{i}X^{b0}_{i+1}....X^{c0}_{k+i}X^{00}_{k+i+1}$, one may consider this region as an effective spin-1 chain with open boundary conditions. The three ladder Hamiltonians of this subsection thus yield exactly soluble  models for such configuration.

An explicit relationship between the original spin-1/2 variables and effective spin-1 operators can also be given by the composite-spin representation \cite{FS}. This approach was used in Ref.~\cite{Albert}.

Considering the ladder with alternating domains of rung triplets and rung singlets, one notices that the couplings $V_{RR}$ and $J_{R}$ are arbitrary, because the Hamiltonian commutes not only with $\sum_{i}X_{i}^{00}$ but also with $X_{i}^{00}$ for any $i$, and therefore the eigenstates of the corresponding ladder Hamiltonian in $4^{N}$-dimensional state space are characterized by the value of the total spin at each rung of the ladder. 
One may study the dependence of the ground state energy as a function of the $J_{R}$, $V_{RR}$ and the total number of rung-singlet bonds. As shown in Ref.~\cite{Albert} depending on these parameters there exist several gapped phases in such a ladder.

\section{Ground state and relation with Matrix-Product Ansatz}
We mention further possible implementations of the dilute algebra used in this analysis.   
The Matrix-Product Ansatz was proven in Ref.~\cite{MPA} to be a valuable technique for the description of the gapped states of spin systems. It is based on a decomposition of the spin Hamiltonian into the sum of projection operators on  plaquette states with fixed angular momentum. There are two plaquette states with total momentum equal to 0 ($j=0$), three triplet states ($j=1$), and one quintuplet ($j=2$). For particular combinations of coupling constants it is possible to find exact ground states. In the current formalism, states with $j=0$ are created by linear combinations of the generators  $p_{i}^{3}$ and $a_{i}^{\dag}$ acting on the product of rung-singlet states, states with $j=1$ are created by the operators $b_{i}\pm b^{\dag}_{i}$, and by the combinations $B_{i}-p_{i}^{0}$ and $p^{1}\pm p^{2}$, and the
state with $j=2$ is given by $6(B_{i}+p^{0}_{i}-4e_{i})$. These states have the form of the Matrix-Product Ansatz \cite{MPA}.  

A natural interpretation also exists for these states as ``words'', the letters of which are elements of the dilute algebra defined by the  operators in Eq.~(\ref{algi}). The relation between exact ground states and words of a (Temperley-Lieb) algebra was noticed first in Ref.~\cite{Words}. It provides a direct algorithm to search for possible exact ground states: one first classifies all possible simple words of the type $K_{i-1}L_{i}M_{i+1}$, where in this analysis the operators $K,L,M$ may be any of $a_{i}, a^{\dag}_{i}, b_{i}, b^{\dag}_{i}, B_{i}$, and $p^{\alpha}_{i}$, and then constructs a product of these elementary words. If the Hamiltonian leaves this product invariant we have obtained an exact eigenstate. In general this procedure can be used for constructing variational wave functions. Examples of such states are given by the following words
\begin{eqnarray}
|\psi_{RS}\rangle\longrightarrow p^{3}_{1}p^{3}_{2}...p^{3}_{N-1} p^{3}_{N}  &  \mbox{and}  & 
|\psi_{FM}\rangle\longrightarrow p^{0}_{1}p^{0}_{2}...p^{0}_{N-1} p^{0}_{N}
\end{eqnarray} 
which correspond to the rung-singlet and ferromagnetic ground states respectively, and  
\begin{eqnarray}
|\psi_{j=0}\rangle = \prod_{i=1}^{N} |\psi^{-}_{s,j=0}\rangle_{i},
\end{eqnarray}
where the local state
\begin{eqnarray}
|\psi^{-}_{s,j=0}\rangle_{i}&=&\frac{1}{(1+3x^{2})^{1/2}}(p^{3}_{i}-x a^{\dag}_{i})|0\rangle
\end{eqnarray} 
produces the variational wave function which interpolates between the rung-singlet state (for $x=0$) and one of the plaquette-singlet states (for $x=1$).

\section{Conclusion}

We have shown that the undeformed dilute two-color braid monoid algebra provides a natural tool for the description of integrable spin-1/2 isotropic ladder models with isotropic nearest and next-nearest-neighbor interactions. The Baxterization of different subsets of generators of the algebra yields corresponding spin Hamiltonians which are non-trivial. Their general feature is the presence of multiple-spin exchange, which may in some cases lead to critical behavior. The ``words'' constructed as a sequence of the elements of the dilute algebra yield variational wave functions of the ground state.

At this point it is worthwhile to compare the present work to related studies of integrable ladder models. In fact, some of the models obtained above are not new. For example the spin-orbital type model of Section 3.2 is known from the work \cite{Wang} on soluble $SU(4)$-invariant spin ladders. The class of models related to the integrable bilinear-biquadratic spin-1 chain (Section 3.7) has also been constructed using the composite-spin representation \cite{FS},\cite{Albert}. A spin Hamiltonian, similar to Eq.(\ref{mag}) has already appeared in Ref.~\cite{AFW}, but the corresponding  $R$-matrix is different. We believe that the connection between these two different $R$-matrices can be understood on the basis of Hecke-soluble models of Ref.~\cite{Alcaraz}. Additional support of this argument is given by Ref.~\cite{BM}. 

In contrast to the models of Sections 3.1, 3.2 and 3.7, which have appeared before in the literature, those of Sections 3.3, 3.5, and 3.6 appear to be new. Recently, Batchelor and coworkers \cite{BGLM} have used operator algebras for constructing soluble $n$-leg spin ladders in terms of  $o(2^{n})$ and $sp(2^{n})$-invariant $R$-matrices based on Temperley-Lieb and Birman-Wenzl-Murakami algebras. But their construction of algebra operators is different from ours. For the two-leg ladder their Temperley-Lieb generator $Q$ has a fugacity factor $-4$ ($Q^{2}=-4Q$), while our generator $E$ has a fugacity factor equal to +4 (see Sections 3.5 and 3.6 ). Correspondingly, the resulting spin Hamiltonians are different as well. We also notice that although the Hamiltonian (\ref{HB}) may be obtained by a $Z_{4}$-transformation from the Hamiltonian of two decoupled spin-1/2 chains, the algebraic structure underlying its integrability was not known before.

Applications of these exactly soluble models to real spin ladders may be made by the use of field-theoretical methods. The exact solutions which describe gapless points in parameter space correspond to certain conformal field theories, WZW models, in the continuum limit. The perturbation of these theories by relevant and marginal operators which correspond to different terms in a real spin Hamiltonian produce in general a renormalization-group flow away from the gapless behavior thus allowing the determination of the properties of the relevant phases in the vicinity of the points of second-order phase transitions. This program is realized in the context of real ladder systems with  cyclic four spin-exchange interactions in Ref.~\cite{GNB}.

\section{Acknowledgement}
We thank F. C. Alcaraz for helpful discussions concerning the model of Eq.~(\ref{mag}) and the referees for useful suggestions. One of the authors (V.G.) is grateful to E. Ivashkevich for highly stimulating discussions. We also thank B. Normand for a careful reading of the manuscript. This work was supported by the Swiss National Science Foundation through grant \# 20-68047.02.

\end{document}